\input harvmac

\Title{\vbox{\hbox{HUTP--97/A013}\hbox{hep-th/9703167}}}
{\vbox{\centerline{Global Anomalies and Geometric Engineering
}
\vskip .1in
\centerline{of Critical Theories in Six Dimensions}}}
\vskip .1in
\centerline{\sl Michael Bershadsky\foot{E-mail:
bershad@string.harvard.edu}
and Cumrun Vafa\foot{E-mail: vafa@string.harvard.edu}}
\vskip .2in
\centerline{\it Lyman Laboratory of Physics, Harvard University}
\centerline{\it Cambridge, MA 02138, USA}


\vskip .2in
\centerline{ABSTRACT}
\vskip .2in
We show the existence of global gauge anomalies in six dimensions
for gauge groups $SU(2),SU(3)$ and $G_2$ coupled
to matter, characterized by an element
of $Z_{12},Z_6$ and $Z_3$ respectively.  Consideration of this anomaly rules
out some of the recently proposed 6 dimensional $N=1$ QFT's which
were conjectured to possess IR fixed point at infinite coupling.
We geometrically engineer essentially all the other models with one
tensor multiplet using F-theory.  In addition we construct 3
infinite series using F-theory geometry which do not have field theory analogs.
All these models in the maximally Higgsed
phase correspond to the strong coupling behaviour of
$E_8\times E_8$ heterotic string compactification on $K3$ with instanton
numbers $(12+n,12-n)$.

\Date{March 97}

The aim of this paper is twofold.  First we discuss
the existence of global gauge anomalies for six dimensional
theories with gauge groups $SU(2),SU(3),G_2$ which puts restrictions
on their matter content, generalizing Witten's
$SU(2)$ anomaly in $d=4$ \ref\witan{E. Witten,  Phys. Lett.
{\bf 117B} (1982) 324.}.
  Next we specialize this to the case
of $N=1$ theories in $d=6$ and find how this restriction
automatically arises from the geometry of F-theory compactifications
on Calabi-Yau threefolds \ref\mv{D. Morrison and C. Vafa, Nucl.Phys. {\bf B473}
(1996) 74-92;
D. Morrison and C. Vafa, Nucl.Phys. {\bf B476} (1996) 437.}.  Moreover
for the case of simple gauge groups we discuss
how the $N=1$ superconformal theories conjectured to exist in
\ref\seib{N. Seiberg, Phys.Lett. {\bf B390}
 (1997) 169.}\ and classified in \ref\swed{U.H. Danielsson,
G. Ferretti, J. Kalkkinen and P. Stjernberg, {\it
Notes on Supersymmetric Gauge Theories in Five and Six Dimensions},
hep-th/9703098.}\
are realized in F-theory, after
we take into account the above global anomaly.  Moreover we construct
three additional infinite series using F-theory which have
no field theory analog
and correspond to a generalization
of the phenomenon of zero size $E_8$ instantons.
   All these theories in the maximally higgsed phase
are equivalent
to strong coupling singularity of heterotic strings for $E_8\times E_8$
gauge group with instanton numbers $(12+n,12-n)$, for $1\leq n \leq 12$.
These critical theories, in the maximally higgsed phases have
already been considered in \ref\wittmax{E. Witten, Mod. Phys.
Lett. {\bf A11} (1996) 2649.}.

\newsec{Global Gauge Anomalies in 6 dimensions}
We consider gauge theories in 6 dimensions with gauge group $G$
with some matter field.
Our considerations will be general and in particular do not
depend on having any supersymmetries.  We will be interested in
fermionic matter in these theories.   Let us
call spinors of opposite chirality $S^{\pm}$.  We will assume
that $(S^+,S^-)$ spinors transform according to some representation
$(R^+,R^-)$
of $G$ respectively.  If $R^+$ is not the same as $R^-$ then the
determinant
of fermions will involve phases and defining them in principle can
lead to local or global anomalies.  We will start from a situation where
local anomalies cancel (via a Green-Schwarz mechanism) and are
interested
in knowing if there are any global gauge anomalies.  Such an anomaly
can arise if the space of gauge transformations is disconnected.

A well known example of this occurs in 4 dimensions for
$SU(2)$ gauge groups.  Since $\pi_4(SU(2))
={\bf Z_2}$ taking the space to be $S^4$
 one can consider a gauge transformation not continuously connected
to identity \witan\ and one can show that if we have an odd
number of fermion doublets
the fermion determinant picks up a minus sign.

One way to prove the $SU(2)$ anomaly, which
will immediately generalize to the case under consideration is to
embed the $SU(2)$ group in a bigger group for which $\pi_4$ is
trivial but which may have local gauge anomalies.
In this way  the global anomaly of $SU(2)$ will
be related to the local anomaly of the higher dimensional group
which is technically easier to deal with.  This is the approach
followed in
\ref\nel{S. Elitzur and V. P. Nair, Nucl. Phys. {\bf B243} (1984) 205.}.  They
consider embedding $SU(2)$ into $SU(3)$
for which $\pi_4(SU(3))=0$.  One starts with $SU(3)$ with a single
fundamantal Weyl fermion and considers the non-trivial $SU(2)$
gauge transformation $g$ on the four dimensional space, which is
taken to be $S^4$.  In the $SU(3)$ embedding this can be
extended to a pure gauge transformations
on the disc $D^5$ whose boundary is $S^4$.  The $SU(3)$ theory
with one Weyl fermion in the fundamental representation is
anomalous
 and the
variation of the phase of the determinants can be expressed
as an integral over $D^5$
$$Z\rightarrow Z {\rm exp ( i \int_{D^5} \gamma_5)}$$
where $\gamma_5(g,A)$ is a specific 5-form \nel.
Moreover the fact that the $SU(2)$ theory has no {\it local}
anomalies implies that $\gamma_5$ vanishes on $S^4=\partial D^5$
and thus the above integral can be viewed as an integral over $S^5$.
One considers the exact sequence
$$\pi_5(SU(3))\rightarrow \pi_5(SU(3)/SU(2)) \rightarrow \pi_4(SU(2))
\rightarrow \pi_4(SU(3))$$
$${\bf Z}\rightarrow {\bf Z}\rightarrow {\bf Z_2}\rightarrow 0,$$
which implies that the basic generator of $\pi_5(SU(3))$
gets mapped to the square of the generator of the $\pi_5(SU(3)/SU(2))$
and that the generator of our anomaly is in the image of the generator of
$\pi_5(SU(3)/SU(2))$.
Using the fact that
 $\gamma_5$ is proportional to
$Tr(g^{-1} dg)^5  +d \eta$ (for some (computable) 4-form $\eta$
which depends on $g$ and the gauge connection $A$)
such that for the basic generator of $\pi_5(SU(3))$ the integral
comes out to be $2\pi$, and the additivity
property of $\int\gamma_5$ one immediately learns that the
global anomaly for $SU(2)$ for a single doublet is $exp(i \pi)$.

In fact exactly the same method works in higher dimensions,
which was one of the main motivations of \nel\ for developing it.
The case of $SU(2)$ in 4 dimensions is part of the general class
considered
in \nel\ where $2n$ dimensional theories
with gauge group $SU(n)$ ($\pi_{2n}(SU(n))={\bf Z}_{n!}$) were
considered.  The same
arguments lead to the conclusion that if we have a theory with
$SU(n)$ gauge theory which is free of local anomalies and which can be
embedded  in $SU(n+1)$
representation $R$, then there is a global discrete anomaly leading
to the phase
\eqn\usf{{\rm exp}({2\pi i A_R\over n!})}
where $A_R$ is defined by
\eqn\aus{Tr_R F^{n+1}=A_R Tr_{\rm f} F^{n+1}+{\rm lower}\ {\rm terms}}
where $Tr_{\rm f}$ refers to trace in the fundamental
representation.  Note that
only $A_R$ mod $n!$ is relevant for the anomaly.
The lower terms in \aus\ are irrelevant as
they give rise to integrals which
vanish on $S^{2n+1}$.
Let us now come to the case of 6 dimensional gauge
theories where the relevant question is whether $\pi_6(G)$ is
non-trivial.  Indeed this is so for the groups $G=SU(2),SU(3),G_2$
where we have
$$\pi_6(SU(2))={\bf Z_{12}}$$
$$\pi_6(SU(3))={\bf Z_6}$$
$$\pi_6(G_2)={\bf Z_3}$$
The fact that there could potentially exist an anomaly
in these cases was anticipated in \ref\salamet{
E. Bergshoeff, T.W. Kephart, A. Salam
and E. Sezgin, Mod. Phys. Lett. {\bf A1} (1986) 267.}\ref\eki{E. Kiritsis,
Phys. Lett. {\bf B178} (1986) 53;
Phys. Lett. {\bf B181} (1986) 416.}.
As noted above the case of $SU(3)$ is a special case
of the situation considered in \nel . However
this case was ruled out in \nel\
because local anomaly cancellations was found to be
too restrictive--this was before the discovery
of the Green-Schwarz anomaly cancellation mechanism.
 Allowing for Green-Schwarz anomaly cancellation
mechanism allows one to construct interesting models
in 6 dimensions which are free of local anomalies but
could potentially
suffer from global anomalies.  The modifications of the computation
in \nel\ due
to the presence of
Green-Schwarz mechanism is relatively straightforward
as we will see below.

We  start
with a theory with no local anomalies, possibly using
the Green-Schwarz mechanism.  We focus on the gauge
groups $SU(2),SU(3)$ and $G_2$ which
could in principle still have ${\bf Z_{12}},{\bf
Z_6},{\bf Z_3}$ global gauge
anomalies.  Let $\alpha$ be
the generator of the global anomaly
 group, with $\alpha^n =1$ for $n=12,6,3$
respectively.  Then for each
representation
$R$ there is an integer $k(R)$ defined mod $n$ where the global
transformation leads to $\alpha ^k(R)$ change in the phase of
the fermion determinant.  The condition for absence of global
gauge anomalies is that
$$\sum_i k(R^+_i)-\sum_j k(R^-_j) =0 \qquad {\rm mod}\quad n.$$
We are interested in finding $k(R)$.

Let us consider the $SU(3)$ case first.  This is essentially
a special case considered in \nel\ as noted above.  The
only novelty here is that we will use Green-Schwarz
mechanism to cancel anomaly and we thus obtain a well defined
perturbative $SU(3)$ theory by including a term $B\wedge tr F^2$
in the action where $B$ is an anti-symmetric tensor field which
transforms under the gauge transformation.
Let us imbed $SU(3)$ in $SU(4)$ where we {\it include} in
the action the $B\wedge  tr F^2$ term needed to cancel local
anomaly for $SU(3)$.  Then the consideration of
\nel\ go through with the only modification being that
the variation of the $B$ field modifies the expression
for anomaly to
$${\rm exp}(i\int_{D_7}\gamma_7)\rightarrow
{\rm exp}(i\int_{D_7}[\gamma_7 -{\rm a}\gamma_3 {\rm tr} F^2]$$
for some fixed constant ${\rm a}$ which makes the integrand vanish on
$\partial D_7=S^6$.  Thus the computation reduces once again
to an integral on $S^7$.  If we consider the modified
7-form
$${\tilde \gamma_7}=\gamma_7 -{\rm a}\gamma_3 {\rm tr} F^2$$
it is again proportional to $Tr (g^{-1}dg)^7+d \eta$ for some
$\eta$ and so the same considerations as in \nel\ go through.

 Let us consider the case of the fundamental
representation for $SU(3)$.   Using \usf\ and \aus\
and since in the standard embedding of $SU(3)$ in $SU(4)$
we have ${\bf 4}\rightarrow{\bf 3}+{\bf 1}$
 we learn that $k_3=1$.  Moreover for the adjoint of $SU(4)$
we get $A=8$.  Since the adjoint of $SU(4)$ decomposes
as ${\bf 8}+{\bf 3}+{\bf \overline 3}+1$ of $SU(3)$
we learn that $k_8+2k_3=8$ mod $6$.  We thus have
$k_8=6=0$ mod $3$.  We can obtain
the results for other representations in a similar way
(see next section); in particular
we find that for the 6 dimensional representation $k_6=-1$.

We now wish to extend this result to the case of $G_2$ and $SU(2)$.
Since the canonical
homomorphism $\pi_6(SU(3))\rightarrow \pi_6(G_2)$ is onto
and using the decompositions $7\rightarrow {\bf 3}+{\overline 3}+1$
and $14\rightarrow 8+3+\overline 3$, we learn that $k_7=1$
and $k_{14}=1$
mod 3
(similarly we can compute $k$ for other representations as will
be discussed in the next section).
For the case of $SU(2)$ and using the fact that the homomorphism
$\pi_6(SU(2))\rightarrow \pi_6(SU(3))$ is onto we learn that
$k_2=2$ and $k_3=8$ mod 12 (one can use the pseudo-reality
to consider the analog of half-hypermultiplets for $SU(2)$ fundamentals
in which case we get a square root of the phase for this anomaly).

Now let us specialize the above results to the case
of $N=1$ gauge theories.  In this case the gluinos
are $S^+$ spinors in the adjoint representation while
the matter fermions is in the
$S^-$ representation.  Restricting our attention to $n_2$
doublets of $SU(2)$, $n_3$ triplets and $n_6$ sextets of $SU(3)$
and $n_7$ fundamentals of $G_2$ we learn that the consistency
condition for absence of global gauge anomalies are
$$SU(2):\qquad 4-n_2 =0 \qquad {\rm mod} \ 6$$
\eqn\nam{SU(3):\qquad n_3-n_6=0 \qquad {\rm mod} \ 6}
$$G_2: \qquad 1-n_7= 0 \qquad {\rm mod} \ 3$$
It is a simple exercise to check that all the known
models of heterotic $E_8\times E_8$ or $SO(32)$
theory compactified on $K3$ (which yield $N=1$ in $d=6$)
 are consistent with the above restrictions imposed by cancellation
of global gauge anomalies.

\newsec{F-theory Realization of Anomaly}

Let us discuss how the F-theory descriptions ``knows'' about the global
anomalies.  The fact that geometry can know about such subtle
field theory facts has already been observed in \ref\dkv{M. Douglas,
S. Katz and C. Vafa, {\it Small Instantons, del Pezzo Surfaces and
Type I' theory}, hep-th/9609071.}\ where a five
dimensional geometry was shown to be ``aware" of a $Z_2$ valued
theta angle coming from $\pi_4(SU(2))$.  Similarly
the geometry was shown to be ``aware" of a $Z_2$ gauge anomaly
considered in five dimensional compactifications in
\ref\recen{K. Intriligator, D. Morrison and N. Seiberg,
{\it Five-Dimensional Supersymmetric Gauge Theories and Degenerations of
Calabi-Yau Spaces }, hep-th/9702198.}.

Here we will consider $N=1$ theories constructed as a local
model in F-theory where the geometry consists of an elliptic
CY 3-fold with a non-compact 2 dimensional base.
Another way of viewing the same theory is as type IIB
compactification on
${\rm 2}_{\bf C}$ -dimensional (complex) base manifold $B$
where the complex coupling $\tau$ can make $SL(2,{\bf Z})$ jumps.
Since the theory is a consistent compactification
one would expect it to be free of both local and global anomalies.
As we shall see the absence of global anomalies follows from
the fact
that the two dimensional
compact part of the D7-brane in $B$ has {\it
integer valued} self-intersection.  Below we will denote
by $D$ this two dimensional subspace of the D7-brane.
In general the D7-brane may have a multiple number of components
in which case we denote them by $D_a$.  Different components
of D7-branes may carry different gauge groups $G_a$.  Here we are
generalizing the notion of D-brane to include more
general groups as allowed by Kodaira classification of singularity
in F-theory \mv\ which would correspond to non-perturbative
enhancements of gauge symmetry from the type IIB perspective.

We first start with reviewing the
Green-Schwarz mechanism in six dimensions (we will follow in this
discussion \ref\sad{V. Sadov, {\it Generalized Green-Schwarz Mechanism
in F-theory}, hep-th/9608008}.).  To cancel
the anomaly in six dimensions via the Green-Schwarz mechanism a certain
8-form
\eqn\form{I^{(8)}=({\rm tr}R^2)^2+{1\over 6}{\rm
tr}R^2\sum_aX^{(2)}_a-{2\over 3}\sum_a X^{(4)}_a+4\sum_{a<b}Y_{ab}}
 should be factorizable $X_8=\Omega_{ij}X^{(4)}_i X^{(4)}_j$. The
polynomials $X_a ^{(n)}$ and $Y_{ab}$ are given as follows
$$X^{(n)}_a={\rm Tr}F^n_a-\sum_R n_{R_a} {\rm tr}_{R_a}F^n_a$$
$$Y_{ab}=\sum_{R_a,R_b'}n_{R_aR'_b}{\rm tr}_{R_a} F^2_a{\rm
tr}_{R'_b}F^2_b~.$$
The ${\rm Tr}$ denotes the trace in the adjoint representation,
$tr_{R_a}$
stands for the trace in the representation $R_a$.
The number of matter multiplets in the representation $R_a$ is denoted by
$n_{R_a}$ and the number of matter multiplets in the
mixed representation --  by  $n_{R_a, R'_{a}}$.  $\Omega$ is
an $n+1\times n+1$ matrix, where $n$ denotes the number
of tensor multiplets.
The anomaly is canceled by adding to action a gauge
non-invariant term  $\int \Omega_{ij} B_i \wedge X^{(4)}_j$.

 It is also convenient to introduce the coefficients $A_{R_a}$ and
$y_{R_a}$, that appear in the decomposition ${\rm
tr}_{R_a}F^4=A_{R_a}{\rm
tr}F^4 + y_{R_a}({\rm tr}F^2)^2$
assuming $R_a$ has two independent order four invariants.
Let $D_a$ denote the components of the D-brane
worldvolume in the base $B$, as discussed above.  Let $K$
denote the canonical divisor of the base (this is a 2-cycle
dual to $-c_1(B)$).  It was shown in \sad\ that the existence
of the Green-Schwarz counterterm can be traced to the
D-brane worldvolume action integrated over $D_a$. Using this relation
it was demonstrated that there is a relation
between geometry of $D_a$ and the representation
they carry (which are of the form $n_{R_a}$ and $n_{R_a, R'_b}$)
\eqn\rrel{\eqalign{
{\rm index}(Ad_a)-\sum_R{\rm index}(R_a)n_{R_a}&=6{(K\cdot D_a)
} \cr
y_{Ad_a}-\sum_R y_{R_a}\,n_{R_a}&=-3{(D_a \cdot D_a)} \cr
A_{Ad_a}-\sum_R A_{R_a}\,n_{R_a}&=0 \cr
\sum_{R,R'}{\rm index}(R_a)\,{\rm index}(R'_b)\,n_{R_aR'_b}&={(D_a\cdot
D_b)}. \cr
}}

In the cases where there are no independent
4-th order Casimirs, as is the case for $SU(2),SU(3),G_2,F_4,SO(8),
E_6, E_7,E_8$, the cancellation of local anomalies can always
be done with Green-Schwarz mechanism for arbitrary representations.
However the second equation above, considering the fact
that $D^2$ is an integer may put some integrality restriction.
For the case of $F_4,SO(8),E_6,E_7,E_8$ the $y_R$ are all divisible
by $3$ and so there is no restriction from the above equation.  However
for $SU(2)$, $SU(3)$ and $G_2$ we do get a restriction.
  We find that
$$SU(2):  16-\sum 2y_{R}n_R=-6 D^2$$
$$SU(3):  18-\sum 2 y_{R}n_R=-6 D^2$$
$$G_2:  10-\sum y_{R}n_R =-3 D^2$$
Note that for the case of $SU(N)$, $2y_R$ is an integer.
For $G_2$, $y_R$ is integer for all $R$.
These conditions show that there is a mod 6 restriction\foot{
For $SU(2)$ if we allowed half-hypermultiplets, this would
be a mod 12 condition.}
in the $SU(2)$ and $SU(3)$ cases and a mod 3 condition in $G_2$ case.
In particular we learn that
$$SU(2):  4-\sum 2y_{R}n_R=0 \qquad mod \ 6$$
$$SU(3):  -\sum 2 y_{R}n_R=0 \qquad mod \ 6 $$
$$G_2:  1-\sum y_{R}n_R =0 \qquad mod \ 3 $$
These are exactly the conditions we found in the
previous section \nam\ if we can identify $2y_R$ with $k(R)$ for
$SU(2)$ and $SU(3)$ and $y_R $ with $k(R)$ for the $G_2$ case.
For the representations considered in the previous
section one can readily check that it agrees and the above
formula generalizes it to arbitrary representations (which
one can also verify using the techniques of the previous section).

\newsec{$N=1$ critical theories in 6 dimensions}

A necessary condition to have non-trivial field theories
in 6 dimensions was studied in \seib.
In particular for the case of one tensor multiplet a complete
classification for the solutions of this condition was given in
\swed.  The structure of these solutions is roughly
as follows:  There are a number of exceptional cases
corresponding to groups of lower ranks, where they
admit finite number of solutions.  Then there are in addition
5 infinite
series (three based on $SU(N)$, one on $SO(N)$ and one on $SP(N)$).
These 5 infinite series form 3-chains (connected by Higgsing
to one another):  In one chain we have $SU(N)$ with $2N$ fundamentals,
in the second chain we have $SP(N)$ with $N+8$ fundamentals and
$SU(N)$ with $N+8$ fundamentals and 1 anti-symmetric representation
and in the third chain we have $SO(N)$ with $N-8$ vectors
and $SU(N)$ with $N-8$ fundamentals and one symmetric representation.

Now we ask which of these theories are realized in string theory.
Taking into account the anomaly we have discovered,
the exceptional cases in \swed\ are indeed all realized
in $E_8\times E_8$ heterotic string compactified
on $K3$.  The infinite series cannot
all be realized in a compact set up (as the rank of the
gauge groups for compactifications are bounded).  However
the first few elements of 4 of the 5 infinite series can be
realized in the $K3$ compactification of  $E_8\times
E_8$ heterotic strings.
The situation is somewhat analogous
to type IIA on $K3$ where the rank of the gauge group is bounded
in the compact case, whereas if we consider non-compact situations,
such as $ALE$ spaces, one can bypass the bound on the rank
of the gauge group. It is not surprising, therefore,
that also here this extension can be done by considering
a local non-compact situation.

The right setup for this construction
turns out to be F-theory on elliptic CY 3-folds with a non-compact
base.
The strong coupling singularity is reached in this setup when
the divisor $D$ of the D7 brane worldvolume vanishes \mv .
Vanishing $D$ implies a local invariance under scale transformation
(zero size remains zero after rescaling the overall
size of the space) which one can thus interpret as IR fixed points.
 The condition for vanishing of $D$ implies a certain restriction on the
divisor $D$, namely the first Chern class of its normal bundle
should be negative, in other words $D^2<0$.
 As it will become clear from the discussion the
self-intersection number $D^2$ is an important
characteristic of the critical theory.

Consider first the groups with vanishing fourth order casimir.
The spectrum of such theories can be read from \rrel:
\eqn\discr{\eqalign{
E_7:&~~ n_{{1 \over 2} 56}=D^2+8=0,1,...7~~~D^2=-1,-2,...-8 \cr
E_6:&~~ n_{27} =D^2+6=0,1,...5~~~D^2 =-1,-2,...-6 \cr
F_4:&~~ n_{26} =D^2+5=0,1,...4~~~D^2 =-1,-2,...-5 \cr
SO(8):&~~ n_{8_{v,s,c}}=D^2+4 =0,1,...3~~~D^2=-1,-2,...-4 \cr
G_2:& ~~n_{7} =3D^2+10=1,4,7~~~D^2=-1,-2,-3 \cr
SU(3):&~~ n_3=6D^2+18=0,6,12,~~n_6=0~~~D^2=-1,-2,-3\cr
SU(2):& ~~n_2=6D^2+16=4,10~~~D^2=-1,-2}}
All the gauge groups that appear in  \discr\ are
subgroups of $E_8$.
The theories with the same $D^2$ are Higgsable into each other.
For example, for $D^2=-2$ we have the following
sequence $E_7 \rightarrow E_6 \rightarrow F_4 \rightarrow SO(8) \rightarrow
G_2 \rightarrow SU(3) \rightarrow SU(2)$. If $D^2=-3$, the same sequence
terminates on $SU(3)$ without any matter.
For $SU(6)$ there is an exceptional
critical theory  with  15 fundamentals and half of the tensor multiplet
(corresponding to $D^2=-1$ \foot{This example was also realized in
\ref\kucha{M. Bershadsky, K. Intriligator, S. Kachru D. R.
Morrison, V. Sadov and S. Kachru, Nucl. Phys. {\bf 481} (1996) 215}.}).

There is also another finite set of examples of $SO(N)$ gauge groups with
$N=7,8,9,...12$ and $N=13$. The first set of examples corresponds to
$D^2=-1,-2,-3,-4$, while the case $N=13$ corresponds to two choices of
$D^2=-2,-4$\foot{The examples with $D^2=-4$ fit into an infinite series
to be discussed in this section.}.
The matter spectrum is given by $N_v=N-4+D^2,~N_s=16(4+D^2)/d_s$, where
$d_s$ is the dimension of spinor representation. $N_s$ counts the
total number of spinors
in cases with
two kinds of spinor representation (like for $SO(12)$).

All these models can be realized  as heterotic compactifications
on $K3$ or as F-theory
compactification with base space
Hirzerbruch surface ${\bf F}_n$ with 7-branes
(where index $n=-D^2$)\mv \kucha
\foot{The case of $SO(13)$ was not considered in \kucha .
It can be realised in
$E_8\times E_8$ heterotic compactification by considering
$SO(13)\times SO(3)\subset SO(16)\subset E_8$ and
choosing the
$SO(3)$ gauge bundle with instanton number $(6+{n })$.The
spectrum of $SO(13)$ theory is $(2n+9){\rm 13}+{(2n+4) \over 4}{\rm 64}$
and $2n=D^2=-4,-2$.}.
The gauge group comes from the singularity in the elliptic fibration
at section at infinity $S_{\infty}$. The structure of elliptic fibration
around $S_{\infty}$ determines the gauge group and the matter spectrum.
The vicinity of the singular locus can
be modeled by a normal bundle to $S_{\infty}$ (the normal bundle in question is
${\cal O}_{{\bf P}^1}(-n)$).
The contraction of section $S_{\infty}$ to
a point corresponds to the strong coupling singularity.

Now we formulate the local model of the IR fixed point for F-theory.
Consider the base of the 3-fold to be the
total space $X^{(-D^2)}$ of the line bundle ${\cal O}_{{\bf P}^1}(D^2)$
(for example for $D^2=-2$ the total space $X$ is the cotangent of $P^1$).
We represent the elliptic fibration in the generalized
Weierstrass form
\eqn\wei{y^2+a_1 xy+a_3 y = x^3+x^2 a_2 + x a_4 +a_6}
where $y$ and $x$ are sections of some line bundles ${\cal L}^3$ and
${\cal L}^2$ on $X$.
The restriction of line bundle ${\cal L}$ on ${\bf P}^1$ is
${\cal O}(2+D^2)$.
Coefficients $a_{i}$ are sections of the bundles
${\cal L}^{i}$. Locally, around the zero section each of the
coefficients $a_i$ has the expansion starting with
$z^{\sigma_i} a_{i,\sigma_i}$.
Taking into account that
$z$ is a section of a line bundle ${\cal O}(D^2)$ we arrive at the conclusion
that each of the coefficients $a_{i, \sigma_i}$ is a
holomorphic section of line bundle ${\cal O}(2i+D^2 i-D^2 \sigma)$.
As we will see the condition of having holomorphic section of
${\cal O}(2i+D^2 i-D^2 \sigma)$ imposes restrictions on
possible values of $D^2$.

The possible types of  Kodaira singularities were analyzed
in \kucha\
using the Tate's algorithm \ref\tate{J. Tate,
{\it Algorithm for Determining the Type of a Singular Fiber in an Elliptic
Pencil} in {\it Modular Functions of one Variable IV},
Lecture Notes in Mathematics, Vol. 476 (Shpringer-Verlag, Berlin, 1975).}.
The algorithm proceeds roughly as follows:
make a change of coordinates $(x,y)$ to put the
singularity in the convenient location, blowup the singularity and
then repeat.
At each stage of this process, after the change of
coordinates has been made, the coefficients in the equation will be
divisible by certain powers of $z$. As a result of applying the
Tate's algorithm one gets the divisibility properties of
$a_i$ encoded in values of $\sigma_i$.

Let us first start with the $SU(2)$ case. In order to get an $I_2$
singularity in the elliptic fibration the coefficients $\sigma_i$
should be equal to
$(\sigma_1,\sigma_2,\sigma_3,\sigma_4,\sigma_6 )=(0,0,1,1,2)$
or $(\sigma_1,\sigma_2,\sigma_3,\sigma_4,\sigma_6 )=(0,1,1,1,2)$.
Both choices lead to the same six dimensional field theories.
The condition of having holomorphic section of  ${\cal O}(2i-ni+\sigma_i n)$
implies that $2i+D^2 i-D^2 \sigma >0$ leading to two choices of $D^2=-1,-2$.
Let us define the local model for $SU(2)$ IR fixed point as
\eqn\wei{y^2+a_1 xy+z a_{3,1} y = x^3+x^2 a_2 + x z a_{4,1} +z^2 a_{6,2}~.}
This construction can be easily generalized to other models as soon as we know
the spectrum of $\sigma_i$ (see Table 2 in \kucha).
For example, for $SU(3)$ gauge group
$(\sigma_1,\sigma_2,\sigma_3,\sigma_4,\sigma_6 )=(0,1,1,2,3)$
and  $(\sigma_1,\sigma_2,\sigma_3,\sigma_4,\sigma_6 )=(1,1,2,2,3)$
for $G_2$.

As already mentioned,
apart from the finite set of exceptional examples that can be realized in
heterotic compactification on $K3$ there are
5 infinite series of IR fixed points that cannot be all realized
in a compact situation.  Also as mentioned above they come
in 3 chains.  The first few elements of these chains
are already realized in the compactification of heterotic string
corresponding to the $D^2=-1,-2,-4$:
The $SU(N)$ series with $2N$ fundamental correspond to $D^2=-2$,
the $SP(N)$ with $N+8$ fundamentals and the $SU(N)$ series
with $N+8$ fundamentals and one antisymmetric tensor correspond
to $D^2=-1$ and the $SO(N)$ with $N-8$ fundamentals appears
at $D^2=-4$.  The series with $SU(N)$ with $N-8$ fundamentals
and the symmetric tensor is not realized perturbatively
(but if it were it should have appeared at $D^2=-4$ because
it is higgsable to the $SO$ series and Higgsing does not
affect the value of $D^2$).

We now ask if we can realize these infinite series.  We consider
the local model in F-theory with the non-compact base being
 the total space of the line bundle
${\cal O}(D^2)$ on ${\bf P}^1$ considered above.
Let us first start with the $SU(N)$ case. The spectrum of
$\sigma$'s can be determined from \kucha.
Namely,
$(\sigma_1,\sigma_2,\sigma_3,\sigma_4,\sigma_6 )=(0,1,k,k,2k)$ for $N=2k$
and
$(\sigma_1,\sigma_2,\sigma_3,\sigma_4,\sigma_6 )=(0,1,k-1,k,2k-1)$ for
$N=2k-1$.
For example, for  even $N$ the  elliptic fibration is given as follows
\eqn\wei{y^2+a_1 xy+z^k a_{3,k} y = x^3+x^2 z a_{2,1} + x z^k a_{4,k}
+z^{2k} a_{6,2k}~,}
and similar expression in case of odd $N$. The coefficients
$a_{i, \sigma_i}$ are
holomorphic sections of ${\cal O}(2i+D^2i-D^2 \sigma_i)$.
The existence of these holomorphic sections implies that
$D^2=-1,-2$. It is clear from the construction that the vanishing
cycle is a smooth sphere. Therefore, for $D^2=-2$ we get a realization of
$SU(N)$
IR fixed points with $2N$ fundamentals.
Similarly, for $D^2=-1$ we obtain a realization of
$SU(N)$ IR fixed point with $(N+8)$ fundamentals and one antisymmetric tensor.

The $Sp(N)$  ($D^2=-1$)   series of  critical theories
are realized in a similar manner. The spectra of $\sigma_i$ are given in
Table 2 of \kucha.   For the case of $SO(N)$, $N>8$, one finds
that as long as $0>D^2 \geq -4$ using Tate's algorithm
one can construct an infinite series of examples.  For the case
of $D^2=-4$ this gives the $SO(N)$ series with $N-8$ fundamentals.
For the other cases when $8\leq N\leq 12$ and $D^2=-1,-2,-3$ we
get the exceptional $SO$ series discussed above.  For $N>13$
and $D^2=-1,-2,-3$ the singularity of the local model
(which would have given spinor in the lower rank $SO$ cases)
prevents an interpretation in terms of matter representations but leads
to a perfectly sensible critical theory with no field theory realization.
Note that in this local model, we can always change the complex
structure, which is the analog of ``higgsing'' and reduce
to the phase where there is no strange singularity (basically
by decreasing the value of $N$).  Moreover it can be shown
that these singularities in complex structure appear at finite
distance in Calabi-Yau moduli space \ref\davem{
D. Morrison, private communication.}.
This situation is similar to the case of small $E_8$ instantons.
Note that a small $E_8$ instanton is the generalization of the matter
in the $56$ of $E_7$ to that of $E_8$ (in the sense that the $56$
of $E_7$ arises from the singularity enhancement to $E_8$ \ref\kv{
S. Katz and C. Vafa, {\it Matter from Geometry}, hep-th/9606086.},
 whereas small $E_8$ instantons arise from singularity enhancement
of $E_8$ to $E_9$ \mv ).
The singularity we encounter here for larger values of $N$ in $SO(N)$
is the generalization of
the singularity which leads to spinor matter for $SO(N)$ for $N \leq 13$
(which comes from enhancement of $SO$ singularity to the exceptional
series of singularities).

As discussed before we have already constructed all the local
$SU$ series allowed
in this local setup.  The only series we have not constructed
using F-theory, which is expected based on field theory
analysis of \swed, is that of
 $SU(N)$ with $(N-8)$
fundamentals and one symmetric tensor.  For this case using the results
of \sad\ one sees that this should be possible and that
the local model is {\it not} a normal bundle over ${\bf P}^1$, but rather a
bundle over $g=1$ surface
with one double point.  It should be
possible to explicitly construct this series in this local setup.

\newsec{D-brane realization of $N=1$ six dimensional SCFTs}

Two of the series we have discussed seem to have a perturbative
D-brane realization. One is the $SU(N)$ series with $N_f=2N$
and the other is $SO(N)$ with $N-8$ fundamentals.

We first start with the $SU(N)$ gauge group.
Consider total space $X^{(2)}$ of the
bundle ${\cal O}_{{\bf P}^1}(-2)$. This space is
the cotangent bundle on ${\bf P}^1$ and
it has a trivial first Chern class.
To realize an $SU(N)$ singularity we  wrap $N$ 7-branes over
the $D={\bf P}^1$ (zero section).
By doing this we immediately introduce some curvature that
has to be canceled by additional 7-branes.  Let us denote the additional
7-branes that intersect $D$ as $\Sigma_i$ ($(D \cdot \Sigma_i)=1$).
Each 7-brane  $\Sigma_i$
intersecting $D$ gives rise to a matter multiplet in the
fundamental representation \ref\sado{M. Bershadsky, V.
Sadov and C. Vafa, Nucl.Phys. {\bf B463} (1996) 398.}.
The
canonical class of $X$ with additional 7-branes is equal to
\eqn\can{12 K=N D+ \sum_{i=1}^{N_f}\Sigma_i+...~,}
where the dots denote the extra 7-branes that do not intersect with $D$.
The condition that the total canonical class is zero implies that
$$12 K\cdot D=0=(ND+\sum_{i=1}^{N_f} \Sigma_i)\cdot D=-2N+N_f$$
which leads to
$N_f=2 N$.

The next case we wish to consider is for $D^2=-4$.
Consider in particular the case with the $SO(8)$  gauge
group without matter.
This case can be viewed as $T^* P^1/Z_2$ orientifold
of type IIB
\ref\sen{A. Sen, Nucl.Phys. {\bf B475} (1996) 562.}
\ref\wittentr{E. Witten, Nucl. Phys. {\bf B471}
 (1996) 195.}.  Putting 8 D7-branes
cancels the charge due to the orientifold and prevents the
coupling from running.  Now if we wish to add $N-8$ more D7-branes
wrapped on the orientifold ${\bf P}^1$, thus getting
gauge group $SO(N)$ we must make sure that the extra
curvature is cancelled by $N-8$ intersecting D7-branes,
just as discussed above for the $SU(N)$ case.  This gives
rise to $SO(N)$ gauge theory with $N-8$ fundamentals.

\newsec{Field Theory versus Geometry}
In this paper we have seen that field theory
consideration is rather powerful in predicting
and classifying possible SCFT's that arise
from string theory compactifications, in that
necessary conditions from field theory
appear to be sufficient.  This was also
found to be the case for certain 5-dimensions
SCFT's considered in \recen.

  However if we are interested in studying
questions beyond just mere classification of SCFT's and in particular
for the properties of conformal theories themselves and
the possible branches of such theories, geometry has the upper hand.
Not only geometry will show whether the necessary
conditions for the existence of SCFT's from the QFT considerations
are sufficient, but it will also point to the existence of fixed
points which have no field theory interpretation
such as the three series we have discovered (see also the example in
\dkv \ref\ms{D. Morrison and N. Seiberg, Nucl.Phys. {\bf B483} (1997) 229.}).
Another aspect involves slight deformations away from SCFT's.
For example for the SCFT's in 5 and 6 dimensions there
are tensionless strings, whose properties are best
understood in the context of geometry.  This
is for example manifest in the constructions of BPS
states for such theories \ref\gan{O. Ganor, Nucl.Phys. {\bf B479}
 (1996) 197.}\ref\kmv{A. Klemm,
P. Mayr and C. Vafa, {\it BPS States of Exceptional Non-Critical Strings},
hep-th/9607139.}.
Another arena where geometry has the upper hand is in the questions
of transitions from one branch to another where
the physical question of transition from one branch to another
is mapped to a concrete geometric question.
We strongly believe that the geometric description is the most
powerful way to think about SCFT's and their properties
and that there are many more physical
properties in store for us which
we need to decode from the geometric realization of SCFT's.

\newsec{Acknowledgments}

We are grateful to A. Johansen, D. Morrison, T. Pantev, V. Sadov
and N. Seiberg  for useful discussions.
This work is partially supported by the
NSF grant PHY-92-18167
and by DOE 1994 OJI award and NSF 1994 NYI award.

\listrefs

\end